\input psfig.tex
\documentstyle{l-aa}
\begin{document}
\thesaurus{03.13.6;11.01.2;11.17.3;13.07.2;}

\title{Is there a correlation between radio and gamma ray luminosities of AGN ?}

\author{A. M\"ucke\inst{1},
M. Pohl\inst{1},
P. Reich\inst{2},
W. Reich\inst{2},
R. Schlick\-eiser\inst{2},
C.E. Fichtel\inst{3},
R.C. Hartman\inst{3},
G. Kanbach\inst{1},
D.A. Kniffen\inst{4},
H.A. Mayer-Hassel\-wander\inst{1},
M. Merck\inst{1},
P.F. Michelson\inst{5},
C. von Montigny\inst{3,6},
and T.D. Willis\inst{4}}

\institute{Max-Planck-Institut f\"ur Extraterrestrische Physik, Postfach 1603, 85740 Garching, Germany
\and
Max-Planck-Institut f\"ur Radioastronomie, Postfach 2024, D-53020 Bonn, Germany
\and
NASA/Goddard Space Flight Center, Greenbelt, MD 20771, USA
\and 
Hampden-Sydney College, P.O.Box 862, Hampden-Sydney, VA 23943, USA
\and
Hansen Experimental Physics Laboratory, Stanford University, Stanford, CA 94305, USA
\and 
NAS/NRC Resident Research Associate}

\offprints{afm@mpe-garching.mpg.de}

\date{Received date; accepted date}

\maketitle
\markboth{M\"ucke et al.: A correlation between radio and $\gamma$-ray luminosities of AGN ?}{ }

\begin{abstract}
The possibility of a correlation
between the radio (cm)- and $\gamma$-ray luminosity of variable AGN seen by
EGRET is investigated. We performed Monte-Carlo simulations
of typical data sets and applied different correlation techniques (partial correlation
analysis, $\chi^2$-test applied on flux-flux relations) in view of a truncation
bias caused by sensitivity limits of the surveys.
For K-corrected flux densities, we find that with the least squares method
only a linear correlation can be recovered.
Partial correlation analysis on the other side provides a robust tool to
detect correlations even in flux-limited samples if intrinsic scatter
does not exceed $\sim 40$ \% of the original $\gamma$-ray luminosity.
The analysis presented in this paper takes into account redshift bias and truncation effects
simultaneously which was never considered in earlier papers.

Applying this analysis to simultaneously observed radio- and $\gamma$-ray
data, no correlation is found. However, an artificial correlation appears
when using
the mean flux. This is probably due to the reduction of the dynamical range in the
flux-flux relation. Furthermore, we show that comparing
the emission in both spectral bands at a high activity state leads to no 
convincing correlation.

In conclusion, we can not confirm a correlation between  
radio and $\gamma$-ray \underline{luminosities} of AGN which is claimed 
in previous works.
\keywords{galaxies: active - quasars: general - gamma rays: observations - Methods: statistical} 
\end{abstract}

\section{Introduction}

Many attempts have been made in the past to investigate correlations between
radio (cm)- and $\gamma$-ray luminosities of AGN (Stecker et al. 1993, Padovani et al. 1993, Salamon \& Stecker 1994). By applying regression and
correlation analysis to the 2.7 GHz- and 5 GHz-luminosity and the $\gamma$-ray emission
in the EGRET energy band, a (nearly) linear correlation was found. The use
of luminosities instead of fluxes, however, always introduces a redshift bias
to the data, since luminosities are strongly correlated with redshift. 
Especially in samples 
which cover a wide range of distance, a correlation will appear in a 
luminosity plot even when there is no correlation in the corresponding
flux densities (Elvis et al. 1978). Feigelson \& Berg (1983) show that,
if there is no intrinsic luminosity-luminosity correlation, no correlation
will appear in the flux-flux relation. On the other hand, though the
redshift dependence can be removed, intrinsic correlations between
luminosites $L_1$ and $L_2$ may be lost in the flux diagrams $f_1$-$f_2$: if $L_{1} \sim L_{2}^B$, 
then $f_{1} \sim f_{2}^{B}z^{2(1-B)}$. For $B \neq 1$ each
value $f_{2}^{B}$ will be multiplied by a 'random' value $z^{2(1-B)}$
causing any intrinsic $L_{1}-L_{2}$ correlation to be smeared out.
The apparent correlation is maintained only when the underlying relationship
is linear (Feigelson \& Berg 1983). 
Furthermore, any observational uncertainties will hamper the finding of
correlations using flux diagrams. It is therefore crucial to estimate the 
influence of the redshift bias
on the correlations 
between the emission
from the two wavebands. Non-parametric partial correlation coefficients 
(Kendall's $\tau$ or the Spearman rank correlation
coefficient $R_s$) have been used to 
deal with this problem (Dondi \& Ghisellini 1995). 
One limitation of these rank correlation 
tests is the failure to take into account possible changes in rank
due to observational uncertainties.\\
Observational flux-limits of the samples are another serious bias which influences any correlation
analysis as they restrict the populated region in the luminosity-luminosity diagram
to a narrow band. Therefore, Feigelson \& Berg (1983) proposed to include all upper limits
to avoid artificial correlations and incorrect conclusions (Schmitt 1985), and suggest the 
use of survival analysis. However, if the censored data points are not distributed 
randomly, but localize a particular area, survival analysis may give misleading
results (Isobe 1989).
Furthermore, this analysis can not account for a bias caused by misidentified
sources or by objects
which are completely lost due to the low sensitivity of the
instrument. Those truncation effects must be seriously considered when EGRET data
are used.\\
Another problem arises when considering the nature of the sources.
Blazars are 
known to be strongly variable
in the $\gamma$-ray as well as
in the radio band on time scales of days to months (von Montigny et al. 1995).
Therefore, simultaneous observations should be the adequate data
for a correlation analysis. However, due to the lack of such data, the mean
(Padovani et al. 1993) or the brightest flux values (Dondi \& Ghisellini 1995)
of the sources have been used in the past. For highly variable sources this choice reduces the
dynamical range in the luminosity-luminosity plot essentially, and can
mimic a correlation.

In this paper, we evaluate the reliability of the partial correlation
analysis in luminosity-luminosity plots as well as $\chi^{2}$-fits in flux-flux 
diagrams for flux-limited truncated samples with strongly variable sources. 
For this purpose we
perform Monte-Carlo simulations of typical data sets with known degrees
of correlation and flux sensitivities, which is described in section 2. 
No censored data are used. 
The simulations provide a useful 
test of how correlation analysis methods deal with such selection effects. 
 
The application of a database of EGRET-blazars in the radio (cm)- and $\gamma$-ray
regime is presented in the last section.

Throughout this paper the Hubble constant $H_0=75$ km \mbox{$s^{-1}$} \mbox{$Mpc^{-1}$}
and the deceleration parameter $q_0 = 0.5$ have been used.

\section{Investigation of correlation techniques}
\subsection{Monte-Carlo simulations and correlation analysis}
We consider a sample of objects with a wide range of distances observed in
two spectral bands. To approximate the real case, we assume a sample
of flat-spectrum radio quasars, 
with strong emission in the radio and $\gamma$-ray regime.
The range of the radio luminosity $L_R$ is chosen so that the simulated
data resemble our database, and is set at $40<\rm{log}{L_R[\mbox{ erg }s^{-1}]}<46$. 
The radio luminosity function $\rho_{radio}$ is taken from Dunlop \& Peacock (1990)
\begin{equation} 
\rho_{radio} \sim [(\frac{L_R}{L_c(z)})^{0.83}+(\frac{L_R}{L_c(z)})^{1.96}]^{-1}
\end{equation}
where $L_c(z)$ is the 'evolving' break luminosity
and was parametrized as $\log{L_c(z)} = 25.26+1.18z-0.28z^2$ with the redshift z.

For the case of a 
$L_{\gamma}-L_R$ correlation the $\gamma$-ray luminosity is calculated
from the relation
\begin{equation}
\log{L_{\gamma}} = A+B\log{L_R}+\epsilon(\sigma) 
\end{equation}
where A and B are constants and free parameters. The term $\epsilon(\sigma)$
is a random noise component following a normal distribution with dispersion 
$\sigma$.

If the radio and $\gamma$-ray luminosities are assumed to be uncorrelated,
the probability of detecting an object with a luminosity L in the $\gamma$-ray regime is calculated
from the luminosity function $\rho_{\gamma}$ of EGRET-blazars (Chiang et al. 1995)
\begin{equation}
\rho_{\gamma} \sim (\frac{L_{\gamma}}{L_B})^{-\gamma_2}\Theta(L_{\gamma}-L_B)
+ (\frac{L_{\gamma}}{L_B})^{-\gamma_1}\Theta(L_B-L_{\gamma})
\end{equation}
where $\gamma_1 = 2.9$, $\gamma_2 = 2.6$, $L_B = 10^{46} \mbox{erg }s^{-1} h^{-2}$
and $h = H_0/100$ km \mbox{$s^{-1}$} \mbox{$Mpc^{-1}$}. A pure luminosity evolution is
incorporated with $L_{\gamma}(z) \sim (1+z)^{2.6}$ (see Chiang et al. 1995). The 
$\gamma$-ray luminosity 
range is again taken from the observation: $44<\log{L_{\gamma}[\mbox{ erg } s^{-1}]}<50$.

The objects are distributed in three dimensional space according to their
evolution properties with redshifts z lying between $0.001<z<2.5$. The differential
redshift distribution $dN(z) \sim \rho_{radio} dV(z)$ ($dV(z)$ = the comoving volume element) 
expected from the radio luminosity function is
\begin{eqnarray}
\frac{dN}{dz} & \sim & (1+z)^{-3.5}[(1+z)-\sqrt{1+z}]^{2}\nonumber\\
& & \hspace*{-1.1cm}[[\frac{\frac{2c}{H_0} (1-\frac{1}{\sqrt{1+z}})S_{lim}}{L_c(z)}]^{0.83}+
[\frac{\frac{2c}{H_0} (1-\frac{1}{\sqrt{1+z}})S_{lim}}{L_c(z)}]^{1.96}]^{-1}
\end{eqnarray}
with $S_{lim}$ the sensitivity limit of the instrument and $c$ the speed of light.
Using the relation
\begin{eqnarray}
L_{band} & = & 0.43 \cdot 10^{62} S_{band} \cdot 
\bar E (1+z)^{\alpha-1}\nonumber\\
    &   & [\frac{(1+z)-\sqrt{1+z}}{H_0}]^2
\end{eqnarray}
with $H_0$ the Hubble constant, $S_{band}$ meaning either the radio flux
 $S_{radio}$ or the $\gamma$-ray flux 
$S_{\gamma}$, 
$\bar E$ the mean photon energy 
and $L_{band}$ the corresponding '$\nu L(\nu)$' luminosity (Weedman 1986), 
we convert the luminosity to the flux. 
The term $(1+z)^{\alpha-1}$ is responsible for the K-correction. 
The average energy spectral indices for the respective spectral bands ($S_{band} \sim \nu^{-\alpha}$) are
$\alpha_{radio}=0$ for the radio band, and $\alpha_{\gamma}=1$
for the $\gamma$-ray regime, where the latter also defines the mean photon
energy of $\bar E = $470 MeV. 
All objects which have fluxes lower than a specified limit are
dropped from the data set. To account for the uneven exposure coverage
of EGRET, we use a
sigmoidal probability distribution for the detection of an object: 
\[ P(S_{lim}) = 1 - [1+\exp{(C(S_{lim}+S_0))}]^{-1} \] 
The parameters C and $S_0$ are chosen to be $C=6.1 (20.5)$ and $S_0=-1.25 (-0.55)$  
for the radio ($\gamma$-ray) band.
This means that the credibility limit in the radio band ranges from
$\sim 0.5 \ldots 2$ Jy which can be considered as a result of an overlay of
several radio catalogs with different credibility limits.
The parameters corresponding to the $\gamma$-ray band are estimated from the
flux threshold map of EGRET made from phase 1 and 2 of the mission.
We created simulated data sets for different correlation slopes (B = 0.8, 1.0, 1.3 and 1.5),
assuming either sensitivity limits as described above or perfectly
sensitive detectors (i.e. flux densities down to 0.1 mJy in the radio band, 
and $10^{-11}$ photon $s^{-1}$ $cm^{-2}$ in the $\gamma$-ray regime are observed).
$cm^{-2}$ in the radio
The simulations were done until each sample contains N = 25 and 12 objects.
which are typical numbers for the simultaneously observed data sets and the flaring 
state samples, respectively.
For each parameter set the whole procedure was then repeated 100 times.

We then analysed the
data sets using  
Spearman's rank order partial correlation
coefficient $R_s$ (Macklin 1982). 
The partial correlation describes the relationship between two 
variables when the third variable is held constant. In this way the strong
redshift dependence of the radio and $\gamma$-ray luminosity can be eliminated.
To date the behaviour of Spearman's correlation coefficient in the presence of
upper limits caused by sensitivity limits of the data sets is not known,
in contrast to Kendall's $\tau$, a further non-parametric correlation coefficient.
The parametric methods are more efficient
if one knows the exact distribution of the data. However, in astronomy we do not know the
exact distribution functions 
and hence the non-parametric
methods are usually preferable.
There has also been some debate whether correlations can be
discovered more secure by comparing fluxes instead of luminosities (
Elvis et al. 1978). 
In order to examine how accurately flux-flux diagrams recover quantitative information
from the data when flux-limits and uncertainties are taken into account, 
we carried out a $\chi^2$-test of a linear relation $y=a+bx$ between the logarithms
of the K-corrected flux densities.
The flux density values were provided
with uncertainties of 4\% and 20\% in the radio and 
$\gamma$-ray band, respectively, which are 
typical values for those spectral bands.
The results are summarized in Table 1.

\begin{table*}
\begin{flushleft}
\begin{tabular}{|c||c||c|c|c||c|c|c|c|}\hline
sample & N & $R_s$ & probability$^{*}$ & corr. ? & $\chi^2_{min}$ & significance$^{**}$ & slope b & corr. ?\\ \hline\hline
uncorrelated, flux-limited & 25 & 0.004 & 0.510 & NO & 223.5 & $2\cdot10^{-10}$ & $-0.086\pm0.112$ & NO\\ \hline
                           & 12 & -0.031 & 0.458 & NO & 96.7 & $5\cdot10^{-5}$ & $-0.048\pm0.095$ & NO\\ \hline
uncorrelated, not flux-limited & 25 & 0.060 & 0.493 & NO & 1070.5 & 0. & $0.013\pm0.030$ & NO\\ \hline
                               & 12 & 0.037 & 0.515 & NO & 451.5 & $7\cdot10^{-16}$ & $-0.004\pm0.047$ & NO\\ \hline
$L_{\gamma} \sim L_{radio}^{0.8}$, flux-limited & 25 & 0.920 & $2\cdot10^{-6}$ & YES & 414.5 & $8\cdot10^{-26}$ & $0.878\pm0.057$ & NO\\ \hline
                                                & 12 & 0.896 & $1\cdot10^{-5}$ & YES & 173.8 & $2\cdot10^{-4}$ & $0.866\pm0.106$ & NO\\ \hline
$L_{\gamma} \sim L_{radio}^{0.8}$, not flux-limited & 25 & 0.980 & $2\cdot10^{-15}$ & YES & 54.3 & 0.089 & $0.816\pm0.029$ & NO\\ \hline
                                                    & 12 & 0.961 & $4\cdot10^{-3}$ & YES & 21.8 & 0.224 & $0.821\pm0.044$ & NO\\ \hline
$L_{\gamma} \sim L_{radio}$, flux-limited & 25 & 0.956 & $2\cdot10^{-10}$ & YES & 7.0 & 0.997 & $1.004\pm0.053$ & YES\\ \hline
                                          & 12 & 0.922 & $3\cdot10^{-3}$ & YES & 3.1 & 0.959 & $1.009\pm0.079$ & YES\\ \hline
$L_{\gamma} \sim L_{radio}$, not flux-limited & 25 & 0.984 & $3\cdot10^{-17}$ & YES & 7.9 & 0.993 & $0.979\pm0.028$ & YES\\ \hline
                                              & 12 & 0.966 & $6\cdot10^{-5}$ & YES & 3.5 & 0.944 & $0.980\pm0.042$ & YES\\ \hline
$L_{\gamma} \sim L_{radio}^{1.3}$, flux-limited & 25 & 0.963 & $5\cdot 10^{-11}$ & YES & 135.5 & 0.023 & $1.056\pm0.058$ & NO\\ \hline
                                                & 12 & 0.935 & $4\cdot10^{-3}$ & YES & 56.9 & 0.105 & $1.047\pm0.091$ & NO\\ \hline
$L_{\gamma} \sim L_{radio}^{1.3}$, not flux-limited & 25 & 0.986 & $3\cdot 10^{-17}$ & YES & 147.5 & 0.010 & $1.200\pm0.029$ & NO\\ \hline
                                                    & 12 & 0.971 & $1\cdot10^{-5}$ & YES & 61.5 & 0.075 & $1.193\pm0.043$ & NO\\ \hline
$L_{\gamma} \sim L_{radio}^{1.5}$, flux-limited & 25 & 0.969 & $4\cdot10^{-9}$ & YES & 237.4 & $2\cdot10^{-9}$ & $1.212\pm0.060$ & NO\\ \hline
                                                & 12 & 0.946 & $2\cdot10^{-4}$ & YES & 99.5 & $4\cdot10^{-3}$ & $1.201\pm0.103$ & NO\\ \hline
$L_{\gamma} \sim L_{radio}^{1.5}$, not flux-limited & 25 & 0.987 & $1\cdot10^{-17}$ & YES & 375.0 & $1\cdot10^{-4}$ & $1.326\pm0.032$ & NO\\ \hline
                                                    & 12 & 0.973 & $8\cdot10^{-6}$ & YES & 164.1 & 0.009 & $1.323\pm0.049$ & NO\\ \hline\hline
\end{tabular}
\end{flushleft}
\caption[tab2]{Results of the correlation analysis of simulated data: partial correlation analysis using Spearman rank order
correlation coefficient $R_s$ (with the probabilities (*) of erroneously rejecting the null hypothesis 
(i.e. no correlation), and a least square analysis with the minimum $\chi^{2}$, the goodness-of-fit probability (**) and the slope b
of the fit as indicated above. 100 trials were carried out with each sample containing 
N = 25 and 12 sources. 
In case of a correlation, a random noise component with $\sigma=0.05$
was chosen. 
The analysis were 
carried out between the logarithms of the luminosities and K-corrected flux densities, 
respectively. For the $\chi^2$-fit, the flux density values were provided
with uncertainties of 4\% and 20\% in the radio and 
$\gamma$-ray band, respectively. A correlation is said to exist for a 
chance probability (*) $<$ 5 \%, and a significance (**) $>$ 68 \% using
the partial correlation analysis and least square method, respectively.}
\end{table*}

\begin{figure}
\vbox{\psfig{figure=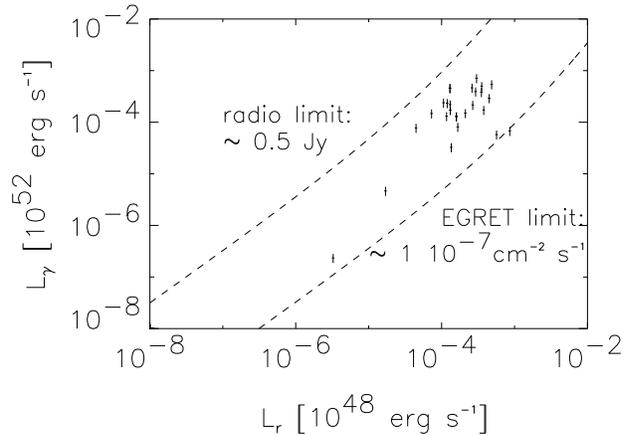,height=6cm,clip=}}
\caption[]{Typical distribution of a {\bf{flux-limited}} sample of objects which possesses {\bf{no correlation}} between
the radio and $\gamma$-ray luminosity. 
The dashed lines follow the sensitivity limits of the sample.}
\end{figure}

\begin{figure}
\vbox{\psfig{figure=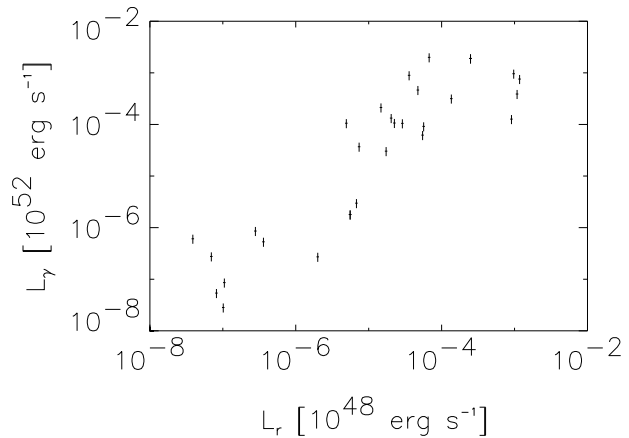,height=6cm,clip=}}
\caption[]{Typical distribution of a {\bf{complete}} sample of objects which possesses {\bf{no correlation}} between
the radio and $\gamma$-ray luminosity.}
\end{figure}

\subsection{Results and discussion}

The results of the partial correlation analysis (see Table 1) show that a correlation
can be recovered at a significance level of $> 99.5 \%$ for both complete and 
flux-limited samples provided that the intrinsic luminosity scatter is not too
large ($\sigma \leq 0.05$). Increasing this noise factor up to 
$\sigma = 0.5$, a correlation can still be detected at an
95 \% level for a complete data set.
The most striking effect of a sensitivity limit is a generally lower correlation coefficient
compared to complete data sets.
High redshift objects can only be detected at high powers.
Hence, sensitivity limits exclude mainly high-redshift objects 
with low luminosities. 
This effect is shown in Fig.1 and Fig.2 where simulated uncorrelated data are
presented for a flux-limited and a complete sample of objects. While
low-luminosity sources are rarely observed in flux-limited data sets, they
are more common in samples which possess no flux-limits.
This causes an overestimation of the
redshift dependence in flux-limited samples. Therefore flux-limits seem to
lower the degree of correlation.

Analysing flux-flux diagrams with the $\chi^2$-tests of a linear relation
can account for observational uncertainties in contrast to a
partial correlation analysis. 
However, only a {\underline{linear}} intrinsic correlation ($L_{\gamma} \sim
L_{radio}$) can be recovered 
with this method no matter whether the sample obeys a sensitivity limit or not.
Except for the case of a linear correlation the derived slopes of the correlations
show misleading trends and the goodness of the $\chi^2$-fit drops down when 
sensitivity limits restrict the flux range.
The low range
of apparent brightness, the 'distance random noise factor $z^{2(1-B)}$', 
observational uncertainties and any intrinsic random noise factor $\epsilon(\sigma)$ 
cause a large scatter which
prevents a reliable goodness-of-fit probability. In addition, we note
that any intrinsic correlation $L_{\gamma} \sim L_R^B$ will be 
randomized by a factor $(1+z)^{\alpha_{\gamma}-1-B(\alpha_{radio}-1)}$ 
($\la 3$ in our case) if the fluxes
are not K-corrected. Especially for samples which cover
a wide range of redshifts but a small range of apparent brightness
it is necessary to use K-corrected fluxes.
Furthermore, when the intrinsic scatter 
causes a standard deviation from the expected luminosity of $\ga 12\%$
($\sigma > 0.05$), we even can not find a linear correlation.
Therefore, an intrinsic correlation could easily disappear using this
method.  
In the case of a correlation with low intrinsic scatter ($\sigma \la 0.05$), the right slopes
are recovered within the parameter uncertainties.

\section{Application to EGRET blazars}
\subsection{The sample}
Our sample consists of 38 identified extragalactic point sources observed by  
EGRET between April 1991 to
September 1993 (phase 1 + 2). EGRET covers the high energy $\gamma$-ray range from about 0.03 to
10 GeV with a field of view of about one-half steradian. 
The typical duration of an observation is two weeks. Only
sources detected $< 25^o$ off-axis and with a significance $> 4\sigma$ 
at least once during the observation run are used in this analysis. 
The flux values
used here are the integral flux ($>$ 100 MeV), and are already published in
Thompson et al. (1995).

Multifrequency radio observations of FSRQ with the 100-m-Effelsberg 
Telescope have been
performed in parallel to the CGRO all-sky survey. All flat spectrum, variable
sources stronger than 1 Jy at 6cm wavelength in the $50^o$ field of view
of EGRET have been observed quasi-simultaneously at 2.8cm, 6cm and 11cm.
Sources, which have been detected by EGRET, were subsequently monitored at
shorter time intervals.
The observational method is described by Reich et al. (1993).

\subsection{Data compilation}
All sources used in this analysis are known to be variable. 
For sources of known redshift, we define $\bar L_{band}$ as the '$\nu L(\nu)$'
luminosity at the indicated frequency.
With a flux S and a redshift $z$, this luminosity $\bar L$ is given 
by equation (5).
For five objects with unknown redshift z, 
we set $z = 1$ which is the mean value
of our sample. For all sources an energy 
spectral index of $\alpha_{radio}=0$ in the radio band and $\alpha_{\gamma}=1$ in the
$\gamma$-ray regime is assumed which are typical values. 

In Fig. 3 the $\gamma$-ray flux is plotted against the simultaneously observed 
2.7 GHz radio flux. The simultaneous observations were done at the Effelsberg
telescope. We denote a pair of radio and $\gamma$-ray 
observations as 'simultaneous' if their observation dates do not differ by 
more than
three weeks. 
Repeated EGRET observations of an AGN are taken as independent sources. 
The corresponding 
luminosity diagram is presented in Fig. 4. 

We also investigated
a possible correlation between the 8 GHz and the 4.8 GHz and the $\gamma$-ray  
luminosities in a flaring state. For the compilation of this data set, we have taken for each 
source the 
highest radio flux found in the literature (Aller et al. (1985), K\"uhr et al. (1981),
 White \& Becker (1982), Wall \& Peacock (1985), Reich et al. (1993), Fiedler et al. (1987),
Seielstad et al. (1983), Waltman et al. 1991, Wright et al. (1990) and the PKSCAT90  
and the maximal $\gamma$-ray flux observed by EGRET. 
In order to ensure that we really
get the peak fluxes, we only use the historical flux values
for $>$ 100 \% variable sources ($ = (S_{max}-S_{min})/S_{min}$ with $S_{min}$ the
lowest detection or upper limit, and $S_{max}$ 
the highest detection ever observed). 

The average
flux values at 4.8 GHz and 8 GHz were determined from the historical minimum
and maximum found in the literature (reference see above). The mean $\gamma$-ray flux was calculated similarly
from the minimum and maximum detection during phase 1 and 2 of the EGRET mission.

\begin{figure}
\vbox{\psfig{figure=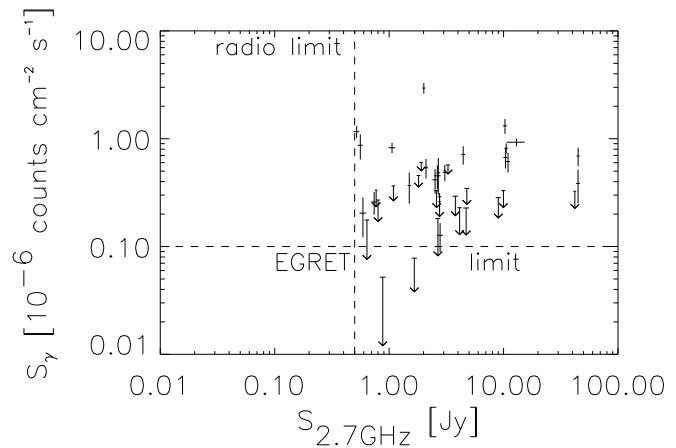,height=6cm,clip=}}
\caption[]{Simultaneously observed $\gamma$-ray flux ($>$ 100 MeV) 
(25 detections and 16 upper limits) versus 2.7 GHz-radio flux density. The dashed line 
follows the sensitivity limits of the sample.}
\end{figure}

\begin{figure}
\vbox{\psfig{figure=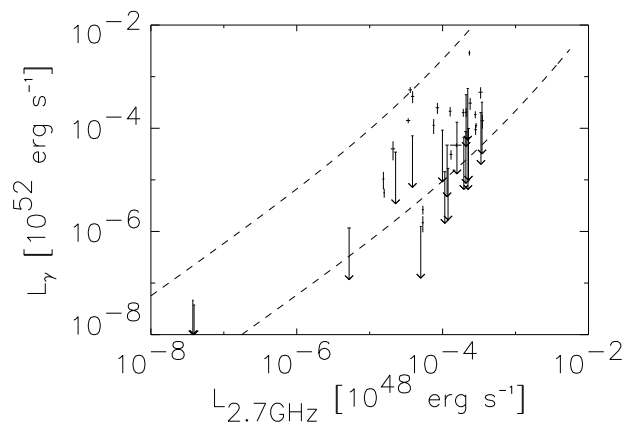,height=6cm,clip=}}
\caption[]{Simultaneously observed $\gamma$-ray luminosity versus 2.7 GHz-radio luminosity. The dashed line 
follows the sensitivity limits of the sample.}
\end{figure}

\begin{table*}
\begin{flushleft}
\begin{tabular}{|c||c||c|c|c||c|c|c|c|}\hline
data & N & corr.coeff. & prob.$^{*}$ & corr. ? & $\chi^2_{min}$ & signif.$^{**}$ & slope b & corr. ?\\ \hline\hline
10 GHz - $\gamma$-ray obs.(simult.det.) & 25 & 0.158 & 0.465 & NO & 352.9 & 0. & $-0.109\pm0.029$ & NO \\ \hline
10 GHz - $\gamma$-ray obs.(simult.det.+UL) & 42 & 0.079 & 0.288 & NO & - & - & - & - \\ \hline
2.7 GHz - $\gamma$-ray obs.(simult.det.) & 22 & 0.046 & 0.751 & NO & 298.5 & 0. & $-0.031\pm0.026$ & NO \\ \hline
2.7 GHz - $\gamma$-ray obs.(simult.det.+UL) & 41 & 0.063 & 0.426 & NO & - & - & - & - \\ \hline
max. 4.8 GHz - $\gamma$-ray obs. & 12 & 0.594 & 0.053 & marg. & 225.9 & $6\cdot10^{-43}$ & $0.450\pm0.029$ & NO\\ \hline
max. 8 GHz - $\gamma$-ray obs. & 11 & 0.363 & 0.314 & NO & 238.0 & 0. & $0.477\pm0.036$ & NO\\ \hline
4.8 GHz - $\gamma$-ray obs. (mean values) & 38 & 0.347 & 0.035 & YES & 159.3 & $2\cdot10^{-17}$ & $-0.225\pm0.041$ & NO\\ \hline
8 GHz - $\gamma$-ray obs. (mean values) & 28 & 0.405 & 0.035 & YES & 144.4 & $2\cdot10^{-18}$ & $0.316\pm0.073$ & NO\\ \hline
\end{tabular}
\end{flushleft}
\caption[tab4]{Results of the correlation analysis of N observed data points: partial correlation analysis using the Spearman rank order
coefficient $R_s$ or Kendall's $\tau$ for the case of data sets where upper limits are
included (with the probabilities (*) of erroneously rejecting the null hypothesis 
(i.e. no correlation)), and a least square analysis with the minimum $\chi^{2}$, the goodness-of-fit probability (**) and the slope b
of the fitted line. The analysis were 
carried out between the logarithms of the luminosities and K-corrected flux densities, 
respectively.}
\end{table*}

\subsection{Results and discussion}
The results of the correlation analysis between different kind of samples
are presented in Table 2. We considered simultaneous detections (including upper limits UL)
at 2.7 GHz and 10 GHz as well as data sets containing the respective highest
flux of each source in the radio and $\gamma$-ray regime. To allow comparison to 
previous investigations, the average flux value of each source in the two 
spectral bands is also correlated.
Analysing K-corrected flux-flux diagrams a positive signal could not be found
for any of the considered samples. We therefore conclude that either an intrinsic
scatter has smeared out a possible correlation or at least a linear correlation
between the radio (cm)- and $\gamma$-ray emission does not exist.

In the last chapter we showed that the Spearman's partial correlation coefficient
can recover intrinsic correlations more secure. Nevertheless, even this method
gives no significant indication for a correlation between simultaneous observations. This is also true when  
upper limits are taken into account and Kendall's partial correlation
coefficient is used. For this method it is possible to state the significance level
even for censored data (Akritas \& Siebert 1996), in contrast to Spearman's $R_s$.

The marginal correlation found between the flaring 4.8 GHz- and $\gamma$-ray
observations looses significance when considering the 8 GHz - $\gamma$-ray 
high-activity state
relation which possess a high chance probability of $ > 30 \%$. 
The flaring 
state observations show only at one radio wavelength a marginal positive result, which we therefore consider as accidental.
Hence, a
convincing correlation between flaring detections does not exist.

Relating the mean 4.8 GHz- as well as the 8 GHz-radio and $\gamma$-ray luminosities gives a 
2 $\sigma$ result for a positive correlation (see Table 2). However,
using averaged flux values in highly variable sources induce a bias which could mimic
a correlation. 
This can be explained as follows:
Suppose, 
a number of variable objects with totally uncorrelated flux values are observed.
The flux at the high activity states of the sources will be found at the upper
right side of the correlation plot while the low activity states are crowded
at the lower left side. After averaging the low and high activity data the
resulting flux values occupy the narrow region in the middle of the 
correlation plot, and will arrange themselves along a straight line
in the luminosity-luminosity-plot. This causes a bias which is not accounted
for by the correlation methods.
Hence, reducing the dynamical range in the flux-flux relation by the averaging procedure 
may cause an artificial 
correlation.   

Our results using Monte-Carlo simulations can not confirm
earlier findings concerning correlations between the radio (cm)- and $\gamma$-ray
emission of blazars (Stecker et al. 1993, Padovani et al. 1993, Dondi \& Ghisellini 1995, Salamon \& Stecker 1994). 
Note, that the correlated quantities they used were not observed simultaneously
although the sources are known to be highly variable. Instead, either the
brightest (Dondi \& Ghisellini 1995) or the mean fluxes (Stecker et al. 1993,
Padovani et al. 1993) were taken. However, the use of average fluxes 
reduces the dynamical range in the flux diagrams and hence may mimic a correlation
(see also Table 2). We conclude that our negative finding of a correlation between 
simultaneously observed sources is likely to be caused by the large scatter of the
data points.
Note also, that truncation effects as well as the influence of the redshift
bias were not considered simultaneously in earlier works. Instead, mostly survival analysis was applied to eliminate biases
caused by the sensitivity limits of the surveys. This method, however, was at
then not able to also deal with a bias caused by the
strong redshift-dependence of the correlated quantities. On the other hand,
Dondi \& Ghisellini (1995) have taken this effect into account by using partial
 correlation analysis, but neglected a bias caused by flux-limits.
We therefore conclude that both, truncation effects as well as redshift bias,
seriously influence the previously performed correlation analysis of radio- and $\gamma$-ray-loud blazars.

\section{Conclusions}
In previous work a correlation between the radio (cm)- and $\gamma$-ray luminosity
of blazars seen by EGRET was claimed. For the correlation analysis 
biases caused by the limited sensitivity of the instruments and the strong
redshift dependence of the luminosities must be considered. While the effects
of flux-limits can be taken into account by including upper limits, partial
correlation analysis deals with the redshift bias. In this paper 
we considered both effects simultaneously in contrast to earlier works. For this purpose we performed
Monte-Carlo simulations of a typical sample of variable blazars with known degree 
of correlation and completeness. We then analyzed the data sets by
partial correlation analysis using the Spearman rank order
correlation coefficient and a least square fit
in flux-flux relations.
The simulations reveal how correlation analysis deals with truncation biases
caused by a limited sensitivity of the instrument. 
While in flux-flux diagrams only linear correlations can be recovered 
(provided that K-corrected fluxes are used) with the least square method, the partial correlation
analysis gives correct results even for flux-limited sample. This is demonstrated
for exponents $B=0.8 \ldots 1.5$ ($L_{\gamma} \sim L_R^B$) with random noise factors
up to an order of magnitude.
Furthermore, in data sets with flux-limits, the partial correlation analysis gives a stronger
redshift dependence of the luminosities compared to complete samples.

Applying the $\chi^2$-method on flux-flux diagrams we found no
 evidence for a
correlation between the simultaneously observed (2.7 GHz and 10 GHz) or 
flaring radio flux and luminosity (4.8 GHz and 8 GHz) and
$\gamma$-ray luminosities above 100 MeV.
The same result is obtained using the partial correlation analysis.
The apparently positive signal between the average
 radio- and
$\gamma$-ray luminosities is caused by the restriction of the dynamical range
in the correlation diagrams.

We conclude that a correlation between the radio (cm)- and $\gamma$-ray 
luminosities of AGN can not be claimed.

Note however, that there is evidence that both radio- and $\gamma$-ray
emissions from blazars are strongly beamed and not isotropic (von Montigny
et al. 1995). The apparent source luminosity is related to the intrinsic
luminosity by a factor which depends on the bulk Lorenz factor of the outmoving
radiation source region within the jet and the angle between the line of sight and the jet axis.
Those quantities are thought to be different for each object and each spectral
band. Therefore, the beaming effect may smear out a possible correlation
between the intrinsic luminosities, or worse, could cause a misleading 
correlation slope. Correlating luminosities which are corrected for beaming
is unfortunately to date not possible since the Doppler factors in the
$\gamma$-ray band are not known from observations.

The fact that we do not find any radio-$\gamma$-ray correlation for both simultaneously
observed luminosities and high activity state observations has several important
implications. 

One may ask whether the lack of a correlation would rule out any of the proposed
$\gamma$-ray emission processes in blazars. Especially the (inhomogeneous)
synchrotron-self-Compton (SSC) models 
(Jones et al. 1974, Marscher 1980, K\"onigl 1981, Marscher \& Gear 1985, 
Ghisellini \& Maraschi 1989, Marscher \& Bloom 1992, Maraschi et al. 1992)
seem to predict a correlation
between the synchrotron emission and the inverse Compton component, as the synchrotron photons are upscattered to $\gamma$-ray energies by the
same beamed relativistic electrons. 
However, there are several reasons why we do not expect to find any relation
between the luminosities in both spectral bands, even if we consider
the SSC mechanism as the main $\gamma$-ray emission process.
First, rapid variability, especially in the $\gamma$-ray regime, causes a
large scatter in the luminosity and flux diagrams, and prevents
recovering the true correlation slope even if a correlation would exist.
Second, different relativistic Doppler factors D for the different objects
induce an additional scatter since
the observer considers different parts of the
spectrum depending on the beaming factor D.
Third, different $\gamma$-ray emission components may contribute differently
for each individual object contaminating the relation with the radio band (i.e. the lack of strong emission lines in BL Lac objects indicates that the 
$\gamma$-ray emission induced by inverse Compton scattering of photons produced 
in the broad line region is small (Sikora et al. 1994)). Note also, that the radio
emission may be influenced by the contribution of a quiescent component
(Valtaoja et al. 1988). 
Fourth, the different variability behaviour in the radio and $\gamma$-ray band
points to a different spatial origin of the two luminosities which excludes a direct
link.
Fifth, there is evidence for a time lag between the radio and $\gamma$-ray
flare (M\"ucke et al. 1996, Valtaoja \& Ter\"asranta 1995, Reich et al. 1993). 
Simultaneously observed luminosities are therefore not expected to obey a 
direct correlation. The same is true for the historical maximum relations
between the radio and $\gamma$-ray regime since those flaring states
must not be necessarily physically connected. 
Instead, the fluence of the two corresponding radio and $\gamma$-ray
flares may be related. However, a single radio outburst can often not
be isolated since radio flares often overlap especially at low cm-frequencies.
Unfortunately, the lack of long-term $\gamma$-ray light curves of blazars
does not enable a meaningful quantitative study at present.
Last, the negative finding of a luminosity correlation could still be compatible 
with a similar energy content of the radio and
$\gamma$-ray producing particles. 
Thus, we do not consider the non-correlation between simultaneously observed
luminosities as strong evidence against the SSC model.

However, the non-correlation yields other astrophysical consequences.
For both, the local luminosity density of blazars and their evolution
properties, there is no hint for a direct link between radio (cm)-
and $\gamma$-ray energies. So, the $\gamma$-ray luminosity function can
not be predicted from a known radio luminosity function and its evolution 
as has been used by several authors 
(Stecker et al. 1993, Salamon \& Stecker 1994, Padovani et al. 1993, Stecker \&
Salamon 1996)
in order to estimate the contribution
of radio-loud AGN to the extragalactic diffuse $\gamma$-ray background.
Instead, we suggest to model the $\gamma$-ray background induced by
unresolved blazars by relating the energy content of particles to the $\gamma$-ray 
emission in AGN
and take into account cosmological distances and the distribution of beaming angles, 
duty factors, etc.. This will be subject of future work.

\end{document}